\documentclass[showpacs,aps,prb,superscriptaddress,floatfix]{revtex4}

\usepackage{graphics}
\usepackage{graphicx}
\begin{document}

\title{Directed diffusion of reconstituting dimers}

\author{Mustansir Barma}

\affiliation{Department of Theoretical Physics, Tata Institute of 
Fundamental Research,\,  Mumbai 400005, India}

\affiliation{Isaac Newton Institute for Mathematical Sciences, 20, 
Clarkson Road, Cambridge CB3 0EH, UK}

\author{Marcelo D. Grynberg}

\affiliation{Departamento de F\'{\i}sica, Universidad Nacional de 
La  Plata,\,  (1900) La Plata, Argentina}

\author{Robin B. Stinchcombe}

\affiliation{Rudolf Peierls Centre for Theoretical  Physics, 
University of Oxford,\, 1 Keble Road, Oxford OX1 3NP, UK}

\affiliation{Isaac Newton Institute for Mathematical Sciences, 20, 
Clarkson Road, Cambridge CB3 0EH, UK}

\begin{abstract}
\vskip 0.2cm 
We discuss dynamical aspects of an asymmetric  version of assisted
diffusion of hard core particles on a ring studied by G. I. Menon {\it
et al.} in J. Stat Phys. {\bf 86}, 1237 (1997). The  asymmetry brings
in phenomena like kinematic waves and effects of the
Kardar-Parisi-Zhang nonlinearity, which combine with the feature of
strongly broken ergodicity, a characteristic of the model. A central
role is played by a single nonlocal invariant, the irreducible string,
whose interplay with the driven motion of reconstituting dimers,
arising from the assisted hopping,  determines the asymptotic dynamics
and scaling  regimes.  These are investigated both analytically and
numerically through sector-dependent mappings to the asymmetric 
simple exclusion process. 
\end{abstract}

%________ Pacs Numbers _______________
\pacs{05.10.Gg,\, 02.50.-r,\,  05.40.-a,\, 02.50.Ey}

\maketitle

%====================
\section{Introduction}
%====================

The issue of universality classes in nonequilibrium statistical systems is
often linked to the existence of conservation laws [\onlinecite{Odor}].  
Normally,  the number of conservation laws is finite, leading to the 
occurrence of dynamically disjoint sectors whose number grows
as a power of the system size.  However, certain  dynamical processes
involving composite objects exhibit strongly broken ergodicity,  with the 
number of disjoint sectors growing exponentially with system size, as a 
result of having an extensive number of conservation laws. 
Examples studied earlier include deposition and evaporation 
($\bullet \bullet \bullet\leftrightarrow \circ \circ \circ$) 
[\onlinecite{BGS,SGB,DB,BD}], and diffusion ($\bullet \bullet \circ 
\leftrightarrow \circ \bullet \bullet$) [\onlinecite{MBD}] of bunches of 
particles. If these moves are  interpreted as involving trimers or dimers 
(in general $k$-mers),  then the $k$-mers in question do not keep their 
identity, and can reconstitute in time. The dynamical moves in
these models do not connect configurations in different sectors,
implying that the steady state is not unique and depends strongly on
the initial condition.  Moreover, the form of the long-time decay of
time-dependent correlations in the steady state varies strongly from one
sector to another.  In one dimension, both the partitioning of phase
space into many sectors and the accompanying dynamical diversity 
could be understood in terms of a nonlocal construct known as the 
Irreducible String (IS) which is an invariant of the motion.  
The IS provided a convenient label for each sector [\onlinecite{DB}].  
Moreover, the position of the elements of the IS were the 
relevant `slow' variables in the problem, and thus governed the long
time dynamics in different sectors. The dynamical diversity found in
different sectors (ranging from different power law decays to
stretched exponentials) could be accounted for in terms of the
differences in the IS from sector to sector [\onlinecite{BD}].
In turn, the IS can be used to construct an extensive number of 
conservation laws, although it turns out that these involve nonlocal 
combinations of the site occupancies [\onlinecite{DB}].

All such studies known to us involve a symmetric movement of the
IS. Here we study the dynamical consequences of an asymmetric
(directed) motion of the IS. This brings in new phenomena associated
with driven diffusive systems, such as kinematic waves and effects of
the Kardar Parisi Zhang (KPZ)/Burgers nonlinearity, present in the
1-$d$ asymmetric simple exclusion process (ASEP) [\onlinecite{RS,GS}]. 
We investigate these effects by studying the directed diffusion of 
reconstituting dimers (DDRD).  
The model resembles the diffusing reconstituting dimer model studied in
Ref. [\onlinecite{MBD}], the only difference being that
we allow for only forward motion of the dimers ($\bullet \bullet
\circ \rightarrow \circ \bullet \bullet$), in contrast to the
two-way motion studied in that paper. The construction of the IS is
the same for symmetric/asymmetric motion, but the dynamics of dimers
(and of the IS) is quite different in the directed case.

It turns out that there is a correspondence between the DDRD and the
well-studied ASEP, though of a generalised sort, leading to extra
features including additional wheeling motion of sites, oscillations 
of correlation functions,  and other new kinetic 
effects. These typically result from the combination of collective
driving and the invariant, but moving, IS. Interestingly,  there is a
very useful correspondence between the DDRD and a 
three-species exclusion process. Further, the problem
of forward-moving non-reconstituting (hard) dimers which keep their
identity is recovered in the DDRD, in a particular sector.  

As in the previous studies of dynamic diversity, Monte Carlo 
(MC) simulation is a key ingredient in trying to understand the
dynamics. In particular we study time-dependent correlation functions
in different sectors and use analytic reductions related to the specific 
form of the IS and the motion of its elements to correlate the diverse 
behaviour seen in the simulations with properties of the ASEP. 
Remarkably close correspondences in nonuniversal as well as 
scaling properties are seen.

The development begins (Section 2) with an introduction to the model  
and then moves on in the same Section to the following sequence 
of emerging topics: equivalence to the three-species process and 
correspondences with the ASEP, whose properties are
summarized.  In Section 3 we study the kinetics
within a particularly simple sector in which the problem is tantamount to
that of non-reconstituting dimers.  New effects such as
wheeling of sites in the equivalent ASEP, resulting  
from the combination of driving and the 
composition of dimers, are shown to have important consequences for
autocorrelation  functions. These show  
early oscillations and later a decay, whose form (exponential or 
power law), is decided by a critical condition related to wheeling and 
kinematic  wave velocities. The long time behaviour shows scaling, and 
universality in the sense of data collapse to ASEP scaling functions,  
but with sector-specific parameters.  
The theoretical and MC results are extended to more general sectors
 in Section 4, via investigations of sublattice currents, sublattice 
current-density relations, and consequent kinematic wave velocities. 
Section 5 is mainly concerned with spatial correlations in a
particular sector, involving MC 
results and their exact analysis via one of the equivalent models. 
Section 6 is a concluding discussion.

%==================================
\section{DDRD Model and Correspondence to ASEP}
%==================================

%----------------------------
%\subsection {Model}
%----------------------------

The DDRD model consists of a ring of $L$ sites, each of which may be
singly occupied (occupation variable $n_i = 1$) or empty ($n_i =
0$). Any of the $2^L$ possible configurations is then an $L$-bit
binary string.  The system evolves stochastically through the move
$110 \rightarrow 011$, which represents the directed diffusion
(rightward hopping), of reconstituting dimers (DDRD). This is
equivalent to stochastic hopping of holes two steps to the left
(i.e. staying on the same sublattice) provided the intervening site is
occupied. The process is the totally asymmetric alternative to the
fully symmetric reconstituting dimer diffusion process (DRD)
[\onlinecite{MBD}]  
which is known to be strongly nonergodic due to the existence of a
conserved string of variables, the IS. All states linked by the dynamic 
process have the same IS, so the phase space divides into many sectors,
$\sim \lambda ^L$ in number, where $\lambda$ is the Golden number 
($\sqrt 5 + 1)/2\,$. All these properties are shared by the asymmetric
generalisations. The IS for a given sector can be obtained from any
state in the sector by deleting from its $L$-bit string any pair of
adjacent $1$'s, and repeating the procedure until no more deletions
are possible (the result is independent of the order of deletion).
For instance, the configuration ${\cal C} \equiv 11101001111010$ 
leads to the IS $10100010$. 

%--------------------------------------------------------
%\subsection {Equivalence of DDRD to 
% exclusion process with three-species, $A, B, C$}
%--------------------------------------------------------

%--------------------------------------------
%\vskip 0.3cm
%{\it \underline {Correspondences}.}
%--------------------------------------------
An equivalent representation of configurations, of the IS, and of the 
process, uses characters $A, B, C$ related to the binary
variables by $A = 11, B = 10, C = 0$ [\onlinecite{MBD}].  With the
periodic boundary conditions used here any DDRD configuration can 
be uniquely decomposed into a configuration of $A$'s, $B$'s, and $C$'s. 
For example, the configuration $\cal C$ of the last paragraph can be 
written as $ABBCAACB$.  Moreover, the totally asymmetric dimer 
hopping move $110 \rightarrow 011$ corresponds to either 
$AB \rightarrow BA$, 
or $AC \rightarrow CA$; the former move involves dimer reconstitution. 
The IS construction now corresponds to deletion of all $A$'s, so the IS 
is a string of only $B$'s and $C$'s, e.g. for the configuration $\cal C$ 
defined above the IS is $BBCCB$.  The absence of any exchange of 
$B$'s and $C$'s in the DDRD process verifies the conservation of the IS.

%-------------------------------------------------------------------------
%\subsection {Correspondence between DDRD and equivalent ASEP}
%-------------------------------------------------------------------------
In terms of the characters $A, B, C$, the DDRD process has an obvious  
and important analogy with the (totally) asymmetric exclusion 
process (ASEP), in which mutually excluding particles hop to nearest 
neighbour vacancies on the right. In this correspondence the dimers 
$A$ play the role of the ASEP particle, and the $B$'s, and $C$'s  of the 
IS correspond to ASEP vacancies.
In order to exploit the correspondence here and later,  we 
use $N_A, N_B, N_C$ to denote the numbers of dimers, 10 pairs, 
and single 0's, respectively. Then the lengths of the DDRD lattice, the IS, 
and the  ASEP lattice are  $L = 2N_A + 2N_B + N_C,\, 
{\cal L} = 2N_B + N_C$, and  $L_X = N_A + N_B + N_C$, respectively.

The particle density in the ASEP is $x = N_A/L_X$, and the  
fraction of zeros of C type is $y = N_C/(N_B + N_C)$. Then, 
in the case of a periodic IS, if the periodic unit [...] of the IS 
contains $n_B$  $B$'s and $n_C$  $C$'s, we have  
$y = n_C/(n_B + n_C)$.   
It is straightforward to show from these definitions that 
$L/L_X = 2 -  y(1 - x)$.  
This ratio provides a metric factor particularly important for converting
known currents, tagged hole velocities,  
and particle densities for the ASEP into corresponding DDRD
quantities. An important point is that the ASEP image of a fixed DDRD
site is not fixed, but moves forward and wheels around the periodic
ring. The causes and consequences of wheeling are discussed in detail
in Section 3. 

%-----------------------------------------------------------------
%\subsection {Correspondence to Zero Range Process}
%----------------------------------------------------------------
We conclude the section by recalling some facts about the ASEP 
that we will need later. 
%==================================
%\section{Kinematic waves and wheeling}
%==================================
%-------------------------------------------------------------------------
%\subsection {Kinetic analogues with, and basic properties of, ASEP}
%-------------------------------------------------------------------------
%The typical snapshot of Fig.\,\ref{snap.ps} displaying the space-time 
%evolution of the DDRD in the null sector $[0]^{3L/5}$, shows collective
%drift  in the  form of shocks with a mean local velocity dependent on the 
%local particle density. This is analogous to kinematic waves  
%encountered in the ASEP (the archetypal collective driven system), 
%and this is one of many parallels encountered in our study. 
%So we digress with a brief reminder of properties of the ASEP 
%(on a ring), and then move on to the 
%generalisations which we expect to see in the different 
%sectors of the DDRD system.  
On a ring, the ASEP has a product-measure uniform steady state 
(SS),  with current  $J_X = x(1-x)$ at ASEP particle density
$x$. Density fluctuations move as a kinematic wave through the system
[\onlinecite{LW,GS}],
with velocity $U = \partial _x \, J_X  = (1-2x)$.
%--------------------------------------------------------------------
% \subsection {scaling and correlation functions of ASEP}
%--------------------------------------------------------------------
Also, in the long time scaling regime, the ASEP density-density 
correlation function $C_X (r,t) \equiv \langle n(r,t)\, n(0,0) \rangle
 - \langle n \rangle ^2\,$ is of the form 
\begin{equation}
\label{scaling}
C_X (r,t)  \propto t^{-2/3} g^{\prime\prime} (s)\;\;\; , \;\;\;
s =  \frac{1}{2}\,\left(J_X\, t^2\,\right)^{-1/3} (\,r - U t\,)\,.
\end{equation}
It is known [Eq.\,(4.8) of [\onlinecite{PS}]]    that in the limit of large $s$,  
$g^{\prime\prime}(s) \sim  \exp \,(\,-\,\mu\, \vert s^3 \vert )$ with 
$\mu \simeq \, -0.295\,$, so the  correlation function is exponentially 
decaying in time at large time $t$. The exception is when the co-moving 
condition  $r =U t$ applies, a special case being the autocorrelation 
function ($r = 0$) when  $x = 1/2$  (making $U = 0$).  In these cases the 
correlation function shows  $t^{-2/3}$ power law decay in time.

In view of the DDRD-ASEP correspondence, one might expect similar 
behaviour for  asymmetrically diffusing dimers,  unlike symmetrically 
diffusing dimers of Ref. [\onlinecite{MBD}], where power law decay
 is always seen.  Detailed 
considerations of the correspondence show that the results are 
broadly in agreement with these expectations, 
but modified by a variety of interesting extra features discussed below. 

%------------------------------------------------------------------
\section{Null Sector}
%------------------------------------------------------------------

We denote a sector whose IS contains no $B$ elements as a null
sector.  In terms of the notation introduced 
earlier, this refers to  $[0]^{{\cal L}}$ with ${\cal L} = N_C$ and $y=1$. 
%Null strings are prominent in this paper, since much of our basic 
%understanding has come from studies of sectors with null strings. 
%We also study sectors where the IS is a periodic 
%repetition of a group [...] of $B$'s, and $C$'s, 
%e.g. $[BBC ... \,] = [10100 ... \,]$.
%-----------------------------------------------------------------------------
%\subsection{Isomorphism to non-reconstituting dimers (in (null) sector)}
%-----------------------------------------------------------------------------
%In the particular case of the null sector $[0]^{{\cal L} = N_C}$ 
In any such sector, the DDRD with an 
even number of particles is isomorphic to directed diffusion 
of nonreconstituting or `hard' dimers. The reason is that every
cluster of 1's must  
contain an even number of particles (otherwise the IS would not be 
all $0$'s), and hence a cluster can be labeled as $DEDEDEDE$, with a
$D$ at the start and $E$ at the end.  Each successive $DE$ pair can be
thought of as a dimer, which retains its identity forever:  the 
elementary move $110 \rightarrow 011$ is $DE\,0 \rightarrow 0DE$.  The
problem of nonreconstituting hard dimers (and more generally $k$-mers)
is of interest in several contexts, and some aspects have been studied
earlier [\onlinecite{RZ,GS2}].  This isomorphism allows us to obtain the 
correlation  function for the reconstituting dimer problem,  in the special 
case of  the null sector, from the derivation given in Section 5 for the 
non-reconstituting case.

We now summarize the results for dynamic properties of interest in
 null sectors. A more detailed account will appear in
 [\onlinecite{SGB2}].  By following the motion of a hole in the DDRD
 and its image in the ASEP, we find  
 that the ratio of the currents  $J_{_{DDRD}}$ (for DDRD), and
$J_X$ is equal to the metric factor  $L_X/L=1/(1+x)$. 
Evidently, we then have  $J_{_{DDRD}} = x(1-x)/(1+x)$.

%---------------------------------------
\vskip 0.3cm 
{\it \underline {Wheeling effect}.}
%----------------------------------------
An important effect shows up in a more detailed consideration of the 
mapping between sites in the ASEP and in the DDRD.  
%-----------------------------------------------------------------------------------
% \subsection {Wheeling effect, and wheeling velocity in the null sector}
%-----------------------------------------------------------------------------------
A fixed site in the DDRD problem corresponds to a moving site in the 
ASEP -- it wheels around the ring at a finite mean velocity. 
This happens because if we follow the ASEP image of a single jump 
$110 \rightarrow 011$ in the DDRD, we find that the image of the 
central site of the triplet advances by one unit, while the images of the
other two sites remain unchanged. In time $t$, the displacement of a 
mapped site is thus $\Delta r (t)=Wt+\phi(t)$ where $W$ is the wheeling 
velocity and $\phi(t)$ is a zero-mean variable which represents the effect 
of stochasticity in the motion.  It can be shown [\onlinecite{SGB2}] 
that in the null  sector, the wheeling velocity is given by 
\begin{equation}
\label{wheel}
W = x\, (1 - x) / (1 + x)\,.
\end{equation}
The wheeling motion of ASEP sites has  been verified 
by direct observation of mapped site motion in MC simulations, which 
shows a small jitter corresponding to $\phi(t)$ around the predicted 
average wheeling velocity.

%---------------------------------------
\vskip 0.3cm 
{\it \underline {Autocorrelation functions}.}
%----------------------------------------
%==================================
%\section{Autocorrelation functions in null sectors}
%==================================
%-------------------------------------------------------------------
% \subsection {Expected generalisations of ASEP scaling, 
% including wheeling velocity of DDRD}
%-------------------------------------------------------------------
The considerations of the previous paragraph have important 
consequences for the DDRD: correlation  
functions are like those of the ASEP 
only if wheeling of sites ($r \rightarrow r+Wt$) is 
allowed for. This suggests that in the long time scaling regime,
 if the effects of jitter arising from the stochastic part $\phi(t)$ 
can be ignored, the density-density correlation function
$C(r,t)$ of the DDRD in  the null sector will be of the form 
\begin{equation}
C(r,t) = t^{-2/3} F \big(\, [\,r + (W-U)t\,] \, t^{-2/3}\,\big)\,.
\end{equation}
For the autocorrelation function ($r = 0$) the argument of $F$ is 
proportional to the difference of the wheeling velocity and 
the kinematic wave speed $U$ of the ASEP, 
$W-U = x (1-x) / (1+x) - (1-2x)\,$. 
Then, if $F$ is similar to the scaling function $g^{\prime\prime}$ of
the ASEP,  
the autocorrelation function $C(r = 0, t) \equiv C(t)\,$ of the null 
sector will decay as  $t^{-2/3} \exp (-\,\kappa\,t)\,$
at large times, where $\kappa$ is a constant proportional to $W-U$. 
On the other hand, in the special 
case where $W-U$ vanishes, late-time slower-than-exponential 
decay should  be expected, that is 
$C(t) \propto t^{-2/3} Y(t)\,$
where $Y$ falls slowly, e.g. approaching a constant. 
%or stretched
%exponential, etc. 
Notice that  $W-U$ vanishes at the `compensating' value  $x_c$ 
of the ASEP particle concentration given by  
\begin{equation}
x_c= \sqrt 2\,-\, 1\,, 
\end{equation}
which would correspond to an IS $[0]^{\cal L}\,$ of length
${\cal L} = L \, (\sqrt 2-1)\,$. 

%-------------------------------------------
\vskip 0.3cm
{\it \underline{Numerical simulations}.}
%-------------------------------------------
To support these expectations, we have conducted extensive
simulations of forward-diffusing, reconstituting dimers in a variety
of situations.  Since autocorrelation functions of interest refer to
spontaneous 
fluctuations in steady state (SS), we avoided the relaxation of the system
by  generating directly SS configurations, thus saving considerable 
CPU time.  In the equivalent ASEP, all configurations in SS are equally 
weighted  under periodic boundary conditions and can be translated 
to DDRD configurations. This enables us to average our measurements 
over a large number of samples to reduce scatter in the data.

In null sectors with IS of length ${\cal L}$, such initial 
conditions are obtained by random deposition of $(L - {\cal L}) /2\,$  
monomers on a ring of  $(L + {\cal L}) /2\,$ sites. Subsequently, 
each monomer is duplicated by adding  another one over an extra 
adjacent location specially  added for that purpose, i.e. $ 1 \to A = 11\,$, 
which leave us with a ring of $L$ sites and $(L - {\cal L}) /2$
randomly distributed dimers.
Notice that this generating procedure is not equivalent to a random
sequential adsorption of dimers in the original system. Had the latter  
method been applied it would necessarily introduce correlations 
between dimers because of shielding effects among them 
[\onlinecite{Evans,MG}].

\begin{figure}
\vskip -5cm
\begin{center}
\includegraphics[scale=0.55]{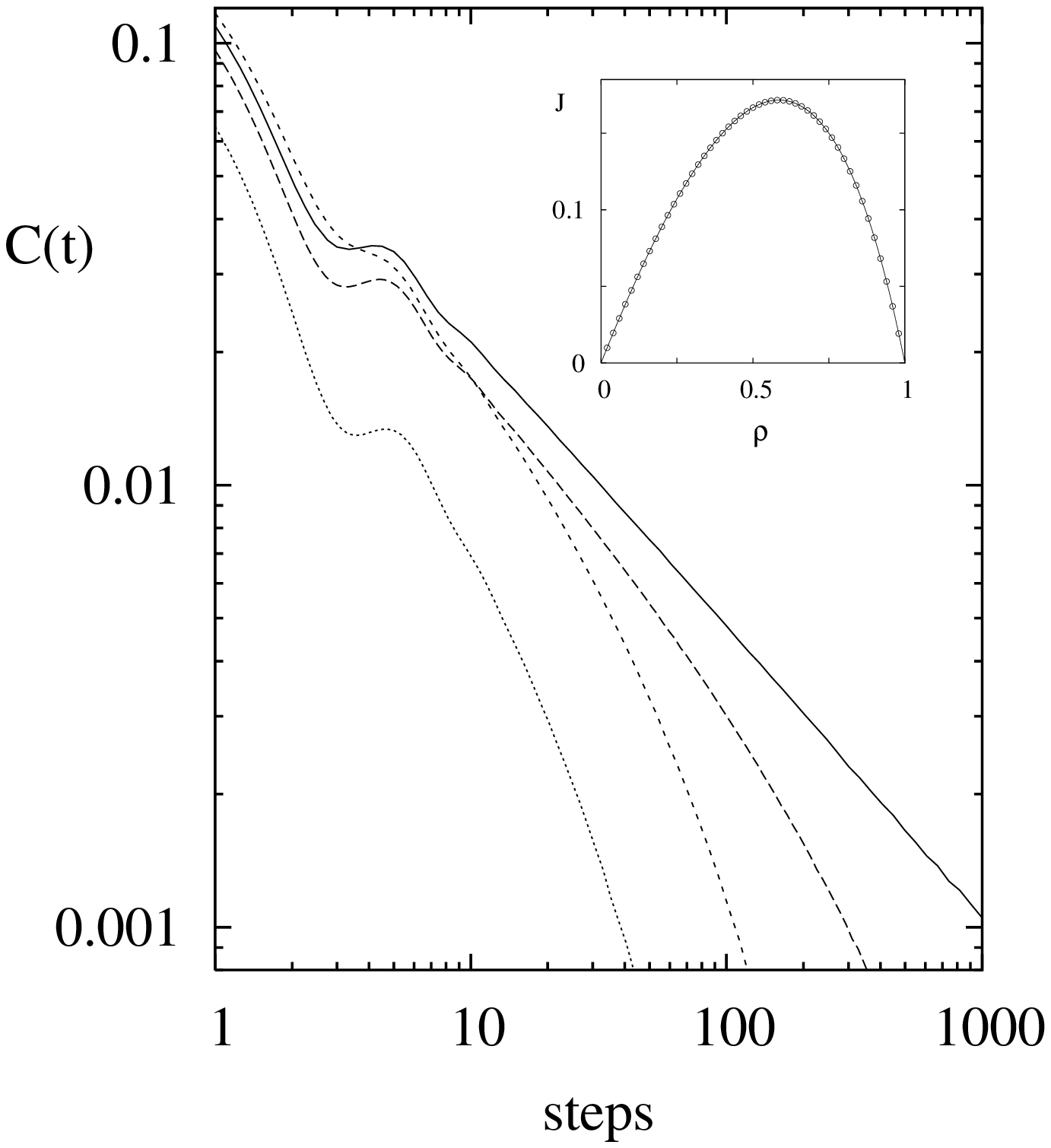}
\vskip -3cm
\caption{Autocorrelation functions of null strings
$[0]^{\cal L}\,$ with different lengths ${\cal L}$. From bottom
to top they refer to ${\cal L }/L = 1/5,\,  3/5, \, 1/3\,$ and  
$\sim \sqrt 2 -1$ (compensating condition $W - U  = 0\,$).
The latter case is consistent with a $t^{-2/3}$ large time decay,
whereas the other situations give rise to exponential decays.
The inset displays steady currents of null sectors with
particle densities $\rho = 1 - {\cal L}/L\,$, closely following the 
analytical current $\rho \,(\rho - 1)/ (\rho -2)$ derived from Section 3,
and reaching a maximum near vanishing wave velocities.}
\label{s0.ps}
\end{center}
\end{figure}

We then evolved our DDRD system using 
the stochastic microscopic rules referred to above. 
After a sequence of $L$ update attempts at random 
locations, the time scale is increased by one unit, i.e. 
$t \to t + 1\,$, irrespective of these attempts being successful.
Also, measurements at shorter (longer) time intervals 
can  also be allowed by updating the system in $M < L$ $\,(M > L)\,$
microsteps, and increasing $t$ as  $t + M/L\,$. Typically, we
considered rings of $10^5$ sites and averaged our
measurements over $2 \times 10^4$ histories starting 
from the independent configurations constructed  above.
%--------------------------------------------------------------------------
% \subsection {Autocorrelation functions for DDRD,  and comparison of 
% autocorrelation scalings at the critical condition in DDRD and ASEP}
%--------------------------------------------------------------------------
In Fig.\,\ref{s0.ps} we show the resulting autocorrelation functions for 
various particle densities $\rho$ in null sectors of length 
${\cal L} = L\,( 1 - \rho)\,$, or equivalently, at different concentrations 
$x = \rho/ (2 - \rho)\,$ of the associated ASEP system. The full line 
corresponds to the compensating  condition $x_c = \sqrt 2 - 1\,$, 
thus making $U-W = 0$, and is the case where slower than exponential 
decay  is seen.  Specifically, it is consistent with a large time behavior
$C(t) \propto t^{-2/3}$,  in turn suggesting that the slow decaying function 
$Y(t)$ referred to after Eq. (3) approaches a constant  at large $t$. 
This is in marked contrast to all other IS fractions  
studied which always give rise to exponential decays, either below or 
above $x_c$. This constitutes a strong confirmation of the theory.

In preparation for the analysis of SS currents in more general situations
(see Section 4), we also estimated numerically these quantities by 
measuring the three-point correlations involved in the 
assisted hopping, namely
\begin{equation}
\label{currents}
J_1 = \frac {2}{L}\, \sum_i \, \langle \, n_{2 i - 1}\,  n_{ 2 i}\,
\overline n_{2 i + 1} \rangle\;, \;\;\;\;
J_2 = \frac {2}{L}\, \sum_i \, \langle \, n_{2 i }\; n_{ 2 i + 1}\,
\overline n_{2 i + 2} \rangle\;,
\end{equation}
on each of sublattices $1$ and $2$, where  $\overline n \equiv 1 - n$
denotes a vacancy. As will be  discussed below,  the compensating   
condition also corresponds to the vanishing  of both current derivatives
with respect to particle densities.  
The inset of Fig.\,\ref{s0.ps} displays the currents of sector $[0]$ 
(both equivalent in this case), as a function of the particle density 
($\rho_1 = \rho_2$), closely following the $J_{_{DDRD}}$ current 
discussed above in this Section. As expected, its extremum occurs 
very near to  $\rho_c = 2 - \sqrt 2$ namely,  close to $x = x_c$.

\begin{figure}
\vskip -4cm
\begin{center}
\includegraphics[scale=0.55]{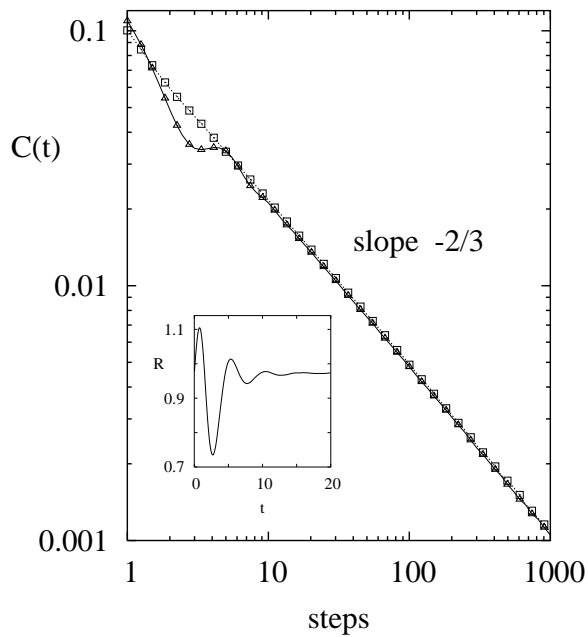}
\vskip -3cm
\caption{Comparison of ASEP (squares) and DDRD (triangles)
autocorrelation functions respectively at $x = 1/2$ and $x = x_c\,$
(compensating conditions in both systems). Above $\sim 20$ MC  steps 
they both decay as $\sim t^{-2/3}$ and with very similar amplitudes.
The short time behavior is shown by the inset  displaying
the ratio $R (t) = C_{_{DDRD}} (t) / C_X (t)\,$ which involves
the wheeling effect referred to in Section 3.} 
\label{asep.ps}
\end{center}
\end{figure}

\begin{figure}
\vskip -4cm
\begin{center}
\includegraphics[scale=0.55]{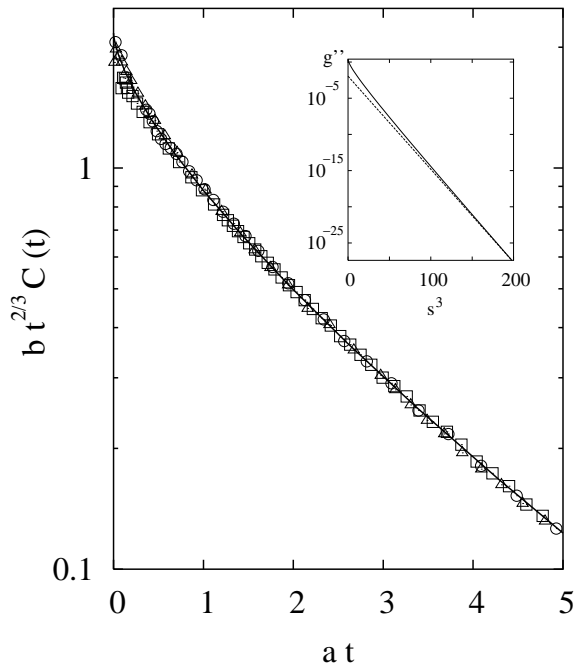}
\vskip -3cm
\caption{Data collapse of autocorrelations in sector $[0]^{\cal L}$ for 
non-critical regimes. They refer respectively to ${\cal L}/L = 3/5$ 
(squares), 1/5 (circles) and, 1/3 (triangles). The curvature of the ASEP 
scaling function   $g''$ (solid line, Ref. [\onlinecite{PS}]), evidences that 
the collapse achieved through the fitting parameters $a,\, b$ is yet very 
far from asymptotia.  The latter is only reached above $s^3 = a\,t 
\gtrsim 200$, as hinted by the actual behavior of $g''$ exhibited in the 
inset. Nevertheless, the resulting values of $a$ and $b$ are 
understandable in  terms of the analysis of $g''$ given in 
Refs. [\onlinecite{PS,FS}] along with the wheeling velocity 
[\,Eq.(\ref{wheel})\,].  For displaying purposes, all early time data 
($a\, t \alt 1/2$) have been pruned.}
\label{colla.ps}
\end{center}
\end{figure}

%----------------------------------------------------
\vskip 0.3cm 
{\it \underline {DDRD and ASEP analogies}.}
%----------------------------------------------------
The above results would indicate that the long time autocorrelations of 
the  DDRD and ASEP systems are similar provided that the
kinematic wave in each system is stationary. Despite the existence 
of jitter between site locations in the two systems, Fig.\,\ref{asep.ps} 
provides a strong indication that this might be the case provided large 
times are considered ($t \gtrsim  20$). On the other hand, the short-time 
differences between both systems  (see inset of Fig.\,\ref{asep.ps} ) 
can be qualitatively understood by combining wheeling with 
the equal-time spatial structure, which, as will be discussed in Section 5, 
has a finite SS correlation length.
%-----------------------------------------------------------
% \subsection {DDRD autocorrelation function Data 
% collapse to ASEP scaling, for non zero $U-W$}
%-----------------------------------------------------------
Moreover, analogues with the ASEP would suggest attempting a 
collapse of all the null sector DDRD autocorrelation data at  
$x \neq x_c$. Fig.\,\ref{colla.ps} shows the data collapse for the DDRD 
autocorrelation $C(t)$ at representative $x$'s in the null sector, as a plot 
of  $\ln [\, b  \,t^{2/3} C(t)\,]\,$  versus  $a\; t$. There is a convincing  
collapse to a line corresponding to $\ln g^{\prime\prime}(s)$ versus 
$s^3$,  where $g^{\prime\prime}$ is the ASEP scaling function 
[Eq.\,(\ref{scaling})\,] in which $s^3 \propto t$ . As for $a$ and $b$, 
they are fitting parameters which depend on $x$, and their numerical 
values [\onlinecite{SGB2}] closely follow those arising from the analysis 
of [\onlinecite{PS,FS}].

%--------------------------------------------------------
% \subsection {Collapse parameters; asymptotic 
% regime; size effects;  initial configurations}
%--------------------------------------------------------
The argument $s$ of $g^{\prime\prime}(s)$ and the overall prefactor
required to give the autocorrelation function for the ASEP are given 
on dimensional grounds as functions of $x$ in [\onlinecite{PS}] and 
[\onlinecite{FS}] (in particular, see Eqs.\,(1.3) to (1.10) in 
[\onlinecite{FS}]\,), and involve $U=(1-2x)$ and  $J_X = x (1-x)$ together 
with $t$ in forms  consistent with the scaling variables given in Section 2.
There are also numerical factors, which differ between the two papers. 
Our data collapse for the DDRD yields values for $a$ which for 
most $x$ are within 1 per cent of the  corresponding theoretical ASEP 
values [\onlinecite{FS}]; likewise  for $b$, except for a
persistent difference of a factor of 2  
(possibly owing to the definition of $g''\,$ in [\onlinecite{FS}]).
Apart from this, the DDRD  autocorrelations correspond in every 
detail to the ASEP ones except at early times.

The data in Fig.\,\ref{colla.ps} do not lie in the asymptotic regime where 
$\ln g^{\prime\prime}(s)$ is linear in $s^3$. This exponential decay
is achieved only for $s^3 \gtrsim 200$, as displayed in upper inset; 
the corresponding times are very large and the autocorrelations too 
small as to render them numerically accessible by direct simulation. 
Size dependences are then noticeable  unless $L^{3/2}$ is large 
compared to such times, so very large system sizes are required in the 
MC. Further, the implied large equilibration times mean that for large 
$L$ it would not be feasible to prepare initial states by relaxation. 
In our simulations, initial states are obtained via the ASEP by exploiting 
its product measure property, as described above.

%==================================
\section{Periodic Irreducible Strings}
%==================================
%-------------------------------------------
% \subsection {Temporal oscillations}
%--------------------------------------------
In non-null sectors, the value of $N_B$ is a measure of reconstitution
that occurs in that sector.  We have examined some sectors with
periodic IS's and find pronounced oscillations in the autocorrelation
functions e.g. Figs. 4 and 5 correspond to sectors $[BCC]^{{\cal
L}/3}$ ($y = 2/3\,$) and $[BBC]^{{\cal L}/3}$ ($y = 1/3\,$), with IS's 
composed of repeating units $BCC$ and $BBC$ respectively.  
The time periods for such periodic sectors are readily calculated, since 
the oscillations are due to the alternations of 1's and 0's arriving
at a site.  

%----------------------------------------------
\vskip 0.3cm
{\it \underline{Steady state construction}.} 
%-----------------------------------------------
In this new scenario the initial steady configurations were
prepared by means of a slight variation of the generating procedure
described for null strings (Section 3).
Consider for instance the sector $[BCC]^{{\cal L}/3}$.  Then, we can
identify two types of  `single site holes'  $B =10$ and $C = 0$, within 
an ASEP ring  of $L/2 + {\cal L}/4$ sites over which  $(L - {\cal L})/2$ 
`monomers' (later on recast as dimers $A = 11$), are randomly 
adsorbed. In going backwards from ASEP to DDRD,  we  first  christen 
the  ASEP vacancies as $B$ or $C$, in the order in which they occur in
the IS. While reconstructing the DDRD configuration, 
we expand out the $B$'s  to be 10's  and let the $C$'s be 0's. 
The resulting configuration is updated
by our  usual stochastic rules. This was carried out on DDRD
rings of  $1.2 \times 10^5$ sites and  averaged typically over 
$3 \times 10^4$ independent histories. 
Particle densities, $\rho_2 = 1 - \frac{{\cal L}}{ L}\,$
and $\rho_1 = 1 - \frac{{\cal L}}{2L}\,$, on even and odd sublattices
are preserved  
throughout [$ x  =  (1 - {\cal L}/L ) / ( 1 + {\cal L}/ 2L )$], 
whereas both autocorrelations and currents were 
separately computed in each sublattice. A similar SS construction
and numerical considerations apply to the $[BBC]^{{\cal L}/3}$ sector, 
with an evolution  which now  takes place in two equivalent sublattices 
of  density  $\rho = 1 - \frac{3 {\cal L}}{5 L}\,$ 
[$x = (1 - {\cal L}/L ) / ( 1 + {\cal L}/ 5L )$].

%--------------------------------------------------------------------
% \subsection {Sublattice currents, kinematic velocities, and 
% autocorrelations in compensating condition $x_c$}
%-------------------------------------------------------------------

%------------------------------------------------------------------
\vskip 0.3cm
{\it \underline {Sublattice currents and wave  velocities}.}
%------------------------------------------------------------------
For sectors such as the above, with periodic IS's, a full discussion
is possible for currents, compensating conditions, and kinematic waves 
with velocities given 
by a density  derivative of the appropriate current in the DDRD model.
%That should reduce  
%to the difference $W-U$ of wheeling and ASEP velocities calculated 
%earlier for the case of the null string.
The compensating condition $W-U=0$ then corresponds to the 
vanishing of the  kinematic wave velocity given by the current derivative
in the DDRD process, at which point we expect slower-than-exponential
decays of autocorrelations. 

%------------------------------------------------------------------------
% \subsection {Analysis for sublattice currents, kinematic wave 
% velocities, and $x_c$, for sectors with periodic IS}
%------------------------------------------------------------------------
In these periodic sectors one can work out 
analytically the current on each sublattice 
using the DDRD-ASEP correspondence given in Section 2, 
and hence obtain the critical $x_c$ and the velocities 
(details can be found in [\onlinecite{SGB2}]).  If $z_1$ and $z_2 =
1-z_1$ are the  
proportions of zeros on sublattices 1,2, then  the current associated 
with the movement of zeros on  each sublattice  is 
\begin{equation}
J_i =2 x \,z_i \,(1-x) / [ 2 - y (1-x) ] \;,\; i = 1,2\,.
\end{equation}
The corresponding DDRD sublattice kinematic wave velocities 
$V_i$ are obtained from $V_i=\partial_{\rho_i} J_i\,$ where $\rho_i$ 
are sublattice particle densities (of 1's) in the DDRD. So $V_1, V_2$ 
vanish at a common value, $x_c$, of the ASEP density $x$  given by 
$\partial_x [x(1-x)/(2 - y(1-x))]  = 0\,$. This is a quadratic equation 
for $x$, whose root lying between 0 and 1 is 
\begin{equation}
x_c = [\sqrt {2(2-y)} - (2-y)]/y\,.
\end{equation}
Thus, $x_c$ is a monotonic decreasing function of $y$ in the range 
$0 \leq y \leq 1$, varying between 0.5 and $\sqrt 2 - 1\,$.  As a check
on the result for $x_c$, for the null sector we have $y=1$, which
implies $x_c= \sqrt 2 -1$.

It is easy to obtain the $V_i$ from the $J_i$ by 
relating the $\rho_i$ to the particle density
$\rho(x)$ in the full DDRD (from both sublattices), 
and thence to $x$ (in terms of which we have $J_i$).  
The result is 
\begin{equation}
V_i = 2- 4x - y (1-x)^ 2\,,
\end{equation}
independent of sublattice label $i$, implying that the two sublattice 
velocities are the same, as well as the $x_c$ at which they vanish.
%----------------------------------------------------------------------
%\subsection {Analytic sublattice current-density relations}
%----------------------------------------------------------------------
For comparison with MC results [\,Eq.(\ref{currents})\,], we can obtain 
similarly,  in any sector with a periodic IS, the current on sublattice  $i$
of the DDRD in terms of the particle density $\rho_i$ on sublattice $i$: 
\begin{equation}
\label{tcurrents}
J_i = (1 - \rho_i)\,  [2 z_i - (2 - y)\,(1 - \rho_i)]  /[2 z_i + y \,(1 -\rho_i)]\,.
\end{equation}

\begin{figure}
\vskip -6cm
\begin{center}
\includegraphics[scale = 0.55]{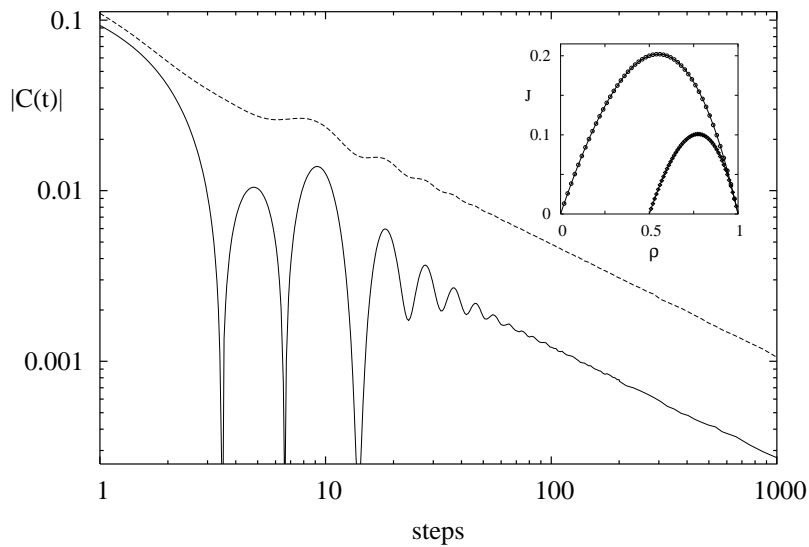}
\vskip -3cm
\caption{Autocorrelation functions of critical $BCC$ sector
($ {\cal L}_c/L \simeq \sqrt 6 - 2\,$), for both even and odd  sublattices
(dashed and solid lines respectively). In both cases, the asymptotic 
behavior is consistent with a $t^{-2/3}\,$ relaxation.
As explained in the text, the early time oscillations are due to 
the periodicity  of the four IS characters [1000].
The inset displays even and odd sublattice currents 
(upper and lower curves respectively), as functions of their
corresponding densities  $\rho_2=1 - \frac{\cal L}{L}$, 
$\rho_1=1 - \frac{\cal L}{2L}$. The data are accurately
described by $J_2 = 2 \rho_2 (1-\rho_2) / (3-\rho_2)\,$ and
 $J_1 = (1 - \rho_1)  (2  \rho_1 - 1) / (2 - \rho_1)\,$, in agreement
with Eq.\,(\ref{tcurrents}). The critical condition of the main panel
corresponds to maximum currents near $\rho_2 =  3 - \sqrt 6\,$, 
and $\rho_1 = 2 - \sqrt{3/2}\,$, namely at vanishing wave velocities.}
\label{bcc.ps}
\end{center} 
\end{figure}

\begin{figure}
\vskip -5cm
\begin{center}
\includegraphics[scale=0.56]{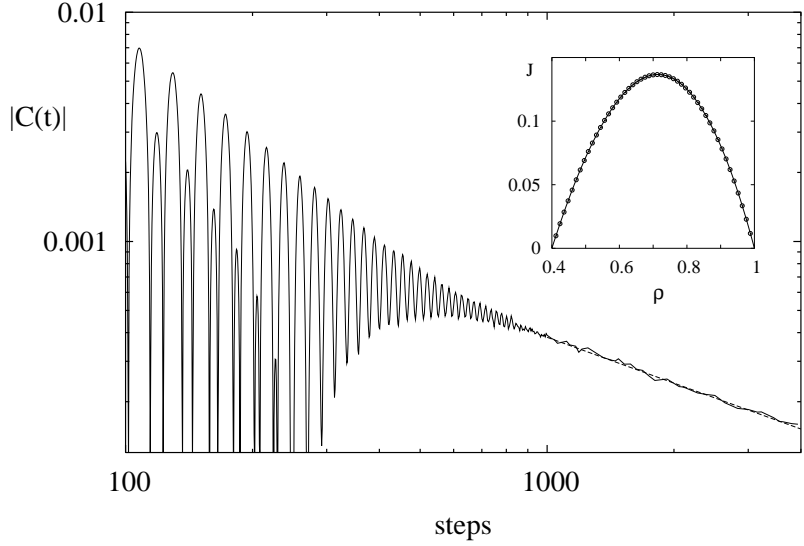}
\vskip -3cm
\caption{Autocorrelations of critical $BBC$ string with length
density ${\cal L}_c /L \simeq \sqrt {30} - 5\,$.
The large time behavior follows closely a $t^{-2/3}$ 
power law decay (denoted by the rightmost lower dashed line).
As in Fig.\,\ref{bcc.ps},  the initial oscillations can be accounted for
by the periodicity of the IS elements [10100]. The inset contains the 
steady currents (equivalent in both sublattices), which follow entirely
their analytical counterparts $J = (\rho - 1)  ( 5 \rho - 2) / (\rho - 4)\,$, 
in Eq.\,(\ref{tcurrents}). The wave velocity vanishes at the current 
maximum, on approaching the main panel regime at 
$\rho = 4 - 3 \sqrt{ 6/5}\,$. }
\label{bbc.ps}
\end{center}
\end{figure}

The analytic results for $[BCC]^{{\cal L}/3}$ and  $[BBC]^{{\cal L}/3}$, 
obtained by inserting their respective $y$ and $z_i$, are the full lines 
in the insets in Figs.\,\ref{bcc.ps} and \ref{bbc.ps} respectively
(skewed parabolae as functions of sublattice densities). Like the null 
string currents discussed in Section 3 (inset of Fig.\,\ref{s0.ps}),
they  agree closely with the data points resulting  from the simulations. 
The sublattice autocorrelation functions shown in the corresponding 
main panels are at the densities of the maxima of  the respective 
sublattice currents, and each show a $t^{-2/3}$ power law long-time 
decay, verifying in turn that they are at the critical density.
Moreover, in analogy to null string cases, it was found that on departing 
from these critical conditions the observed power law decay 
changes abruptly to exponential. 

%==================================
\section{Spatial correlations in null sectors}
%==================================

\begin{figure}
\begin{center}
\includegraphics[scale=0.55]{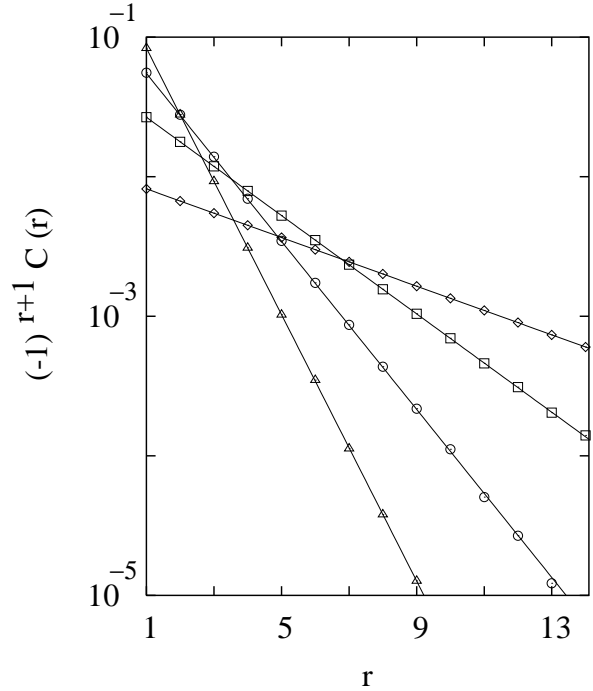}
\caption{Spatial pair correlations of sector $[0]^{\cal L}$
for ${\cal L}/ L = $ 1/2 (triangles), 1/3 (circles), 1/5 (squares), 
and 1/10 (rhomboids). The solid lines are fitted with 
slopes (inverse correlation length) 
$\xi^{-1} = \ln \left(\, \frac{1 + {\cal L}/L}{ 1 - {\cal L}/L} \,\right)$,
in agreement with Eqs.\,(\ref{spatial}) and (\ref{correlation}).
The actual oscillations of $C(r)\,$ also follow the behavior predicted
in Section 5. }
\label{ss.ps}
\end{center}
\end{figure}

%-----------------------------------------------------------------
% \subsection {Steady state spatial correlation function 
% for null string sectors}
%-----------------------------------------------------------------
In the null sector $[0]^{{\cal L } = N_C}$ with an even number of 
particles, the correlation function in the dimer problem can 
be evaluated exactly (below) using the isomorphism to 
non-reconstituting dimers together with the $DE$ representation
introduced in Section 3. Since the presence of a $D$ particle 
at site $i$ implies and is implied by the presence of an $E$ 
particle at site $(i+1)$, the (unsubtracted) particle-particle 
correlation function can be reduced to 
\begin{equation}
\label{spatial}
\langle  n(i)\, n(i+r) \rangle  = 2 {\cal E}(r) + {\cal E}(r-1) +
{\cal E}(r+1) \,,
\end{equation}
where ${\cal E}(r)$ is the (unsubtracted) correlation function 
obtained from all configurations 
having an $E$ particle at each of sites $i$ and $i + r$.
The weight of each such configuration can be 
found by mapping it to the
corresponding configuration in the equivalent ASEP.
Allowing for the weights of the $E$'s at  $i$ and $i + r$
and those of $m$ dimers and $(r-2-2m)$ holes between, and for 
multiplicities, and summing over $m$  (from 0 to the integer 
part of $(r-2)/2\,$) provides the required ${\cal E}(r)$. 
After reductions it becomes
\begin{equation}
\label{correlation}
{\cal E}(r) = [x^2 /(1+x)^2] \, [1 - (-x)^{r-1}]\,,
\end{equation}
(for details, see [\onlinecite{SGB2}]). Inserting into Eq.(\ref{spatial}) 
above,  and subtracting $\langle n \rangle^2 = [2x/(1+x)\,]^2$
provides the subtracted pair correlation function $C(r)$.
This is oscillatory, and has a decaying envelope with correlation 
length  $\xi =[\ln (1/x)]^{-1}\,$.
%--------------------------------------------------------------------
% \subsection {MC spatial correlation function and lengths}
%--------------------------------------------------------------------
Fig.\,\ref{ss.ps} shows MC results  for the  (subtracted) 
spatial correlations $C(r)$, in the null sector for various $x$. The actual 
$C(r)$'s show the predicted oscillation $(-1)^r$, and the correlation 
lengths agree closely with the theory.

%========================
\section{Concluding discussion}
%========================

%-------------------------------
% \subsection {Summary}
%-------------------------------
This investigation of the simplest model combining strongly broken 
ergodicity with driving has uncovered a wealth of properties, some 
expected and others not foreseen. In particular  
the model shows an interesting interplay between the collective 
(and scaling) kinetics of the basic driven model (the ASEP) 
and effects of the invariant IS which characterises 
the sector in the DDRD. Monte Carlo simulations shows man of these 
features in striking fashion. 

Despite the richness and complexity of the combined model, 
much has been understood analytically, for sectors with null 
or periodic IS's. The scaling behaviour seen so far appears to be 
in the universality  class of the ASEP. The relationships of
nonuniversal variables, like currents,  
densities (including critical densities), and velocities etc., have been 
largely understood.

%---------------------------------------------------------
% \subsection {Further questions/generalisations}
%---------------------------------------------------------
There are however certain important generalisations which have not
 been discussed here, which we now briefly mention: 
additional special sectors, alternative boundary conditions, 
and $k$-mer generalisations.

We can expect, following the discussion of Ref. [\onlinecite{MBD}] 
for the DRD, new effects in sectors having IS's with various types of 
structural  correlation. Provided critical conditions ($x_c$'s) exist in
which  autocorrelations show power law time decay, the power law for
such sectors need no longer be related simply to the ASEP exponents. 
One would then have different decay patterns, according to 
sector (`dynamic diversity'). The combination of the methods 
developed here with those of Ref. [\onlinecite{MBD}] might well allow 
analytic treatment of such effects.

%other boundary conditions

Changing from periodic boundary conditions to open ones with 
boundary injection could strongly alter the behaviour. For the ASEP, 
that change produces  non-trivial spatial correlations and dynamics 
and, most important of all, a steady state nonequilibrium phase 
transition. Density profiles, correlations, and dynamics differ in the 
different phases. We would expect all these features to occur in 
the corresponding boundary-driven version of the DDRD, in cases 
(like injection/ejection of dimers) so long as the injection and ejection 
processes keep the IS in tact. Again, the detailed properties 
would vary from sector to sector, and the methods used here should 
prove useful in this generalised situation.

Finally, generalisations from reconstituting dimers to $k$-mers,
deserve consideration. As with dimers, the problem of
nonreconstituting $k$-mers which is of current interest
[\onlinecite{RZ,GS2}] is contained in null sectors of the general
problem. The sector-wise mapping to the ASEP still holds, but the
number of sectors is much larger for larger $k$, as multiplicities in
the types of ASEP holes increase with $k$. The sector-wise study of
the dynamics remains to be explored systematically.

%========================
\section*{Acknowledgements}
%========================

We are grateful to  D. Dhar, G. Sch\"utz, R. Zia, C. Godr\`eche, 
P. Ferrari, H. Spohn, S. Majumdar,  S. Chatterjee, M. Evans, and 
M. Moore for helpful observations and correspondence.
MB's stay at the Newton Institute was supported through 
EPSRC Grant 531174.  MDG acknowledges support of CONICET, 
Argentina,  under Grants  PIP 5037 and PICT ANCYPT 20350.
The work of RBS  was supported by EPSRC under the Oxford 
Condensed Matter Theory Grants GR/R83712/01 and GR/M04426.

%============  References ===================

%============= Figures and Captions ===============
%-------- Fig. 1:  Autocorrelations of Sector [0] + inset -------------
\newpage
  
%-------  Fig. 2:  Comparison between DDRD and ASEP + inset ---------

%----------  Fig. 3:  Data collapse in Sector [0] + inset ------------

%-------- Fig. 4:  Autocorrelations in [BCC] + inset --------------

%-------- Fig. 5:  Autocorrelations in [BBC] + inset --------------

%--------- Fig. 6:  Spatial correlations in Sector [0] ----------------

\end{document}